\documentclass[traditabstract,letter]{aa}
\usepackage{epsfig}
\usepackage{times}
\begin{document}
\title{Stray light correction and contrast analysis of Hinode broad band images}
\author{
S. K. Mathew\inst{1,2}
\and V. Zakharov  \inst{2}
\and S. K. Solanki\inst{2,3}
}
\institute{Udaipur Solar Observatory, P. O. Box 198, Udaipur - 313004, India\\
\email {shibu@prl.res.in}
\and Max-Planck-Institut f\"{u}r Sonnensystemforschung (MPS), Max-Planck-Str. 2,  
37191 Katlenburg-Lindau, Germany\\
\email{zakharov@mps.mpg.de, solanki@mps.mpg.de}
\and School of Space Research, Kyung Hee University, Yongin, Gyeonggi 446-701, Korea\\
}
\offprints{Shibu K. Mathew\\
\email{shibu@prl.res.in}}
   \date{Received ..................; accepted .....................}
    \abstract{
The contrasts of features in the quiet Sun are studied using 
filtergrams recorded  by the Broad-band Filter Imager on the Hinode/Solar Optical Telescope. 
In a first step, the scattered light originating in the instrument is modeled using Mercury transit data.
Combinations of four two-dimensional Gaussians with different widths and weights were 
employed to retrieve the  point-spread functions (PSF) of the instrument at different 
wavelengths, which also describe instrumental scattered light. 
The parameters of PSFs at different wavelengths are tabulated. The observed images were then
deconvolved using the PSFs. The corrected images were used to obtain contrasts of features 
such as bright points and granulation in different wavelength bands. After correction, 
{\it rms} contrasts of the granulation of between 0.11 (at 668 nm) and 0.22 (at 388 nm) are obtained.
Similarly, bright point contrasts ranging from 0.07 (at 668 nm) to 0.78 (at 388 nm) are found, which are
a factor of 1.8 to 2.8 higher than those obtained before PSF deconvolution. The mean contrast of the bright 
points is  found to be somewhat higher in the CN-band than in the G-band, which confirms 
theoretical predictions.  
}
\authorrunning{Mathew, S. K. et al.}
\titlerunning{Stray light correction and contrast analysis}
\maketitle
\section{Introduction}
The contrast  of granulation and of magnetic features is an important diagnostic 
of their thermal structure and provides insight into the energy transport mechanisms 
acting in them. The contrast of granulation, bright points and other small scale features 
is influenced by the point spread function (PSF), the width of 
whose core is a measure of the spatial resolution, while the strength of the wings is 
determined by the amount of light scattered within the instrument (or in the atmosphere,
if present). In particular, the scattered light strongly reduces the contrast. Bright 
points, which are smaller in size than the granules, are more strongly affected. If the PSF is 
known, then it can be used to deconvolve the observed image and thus to approximately retrieve 
the original intensities.

Contrasts measured with the Hinode Solar Optical Telescope 
(SOT, Tsuneta et al. \cite{tsuneta}, 
Suematsu et al. \cite{suematsu}, Kosugi et al. \cite{kosugi})  
are of particular interest due to the combination of high spatial resolution and the absence 
of seeing, leading to almost constant observing conditions. The value of
the Hinode SOT observations would be further enhanced if the PSF could be determined and 
compensated for (e.g. as done by Mathew, et al. \cite{mathew} for MDI continuum
images).

The PSF of the Hinode Spectro Polarimeter (SP) was determined by Danilovic et
al. (\cite{danilovic}) by modeling the SOT/SP optical system using the ZEMAX optical
design software. After convolving solar granulation data from MHD simulations
with the computed PSF, they found that the {\it rms} contrast of the simulated
granulation matches closely the observations from SOT/SP. 
Wedemeyer-B\"{o}hm (\cite{bohm}) used Mercury transit and eclipse images to 
obtain the PSF of the Hinode/SOT/Broadband Filter Imager (BFI) instrument, 
but did not apply the obtained PSF to deducing the true contrast in the BFI 
wavelength bands. 
\begin{figure}
\centerline{{\epsfig{file=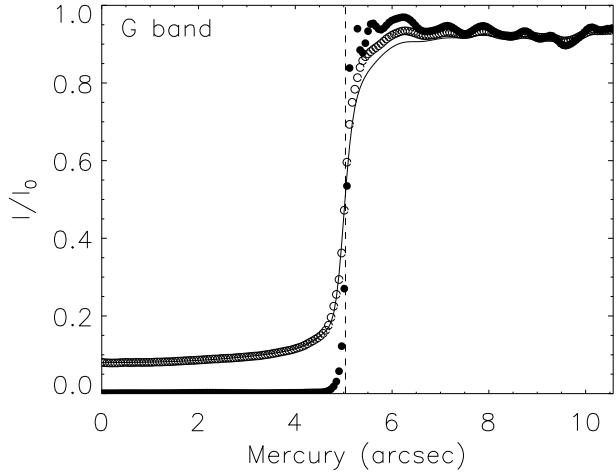,width=9.5cm}}}
\caption{The fit to the radially averaged Mercury intensity profile. Open circles
outline the observed intensity profile, the solid line represents the fit to the
observed profile and filled circles follow the intensity profile after
correction. The vertical dashed line marks Mercury's limb. The ripples in the
intensity profile outside Mercury's image are mainly due to the average
intensity variations in the granules.}
\end{figure}
In this paper we obtain the PSF of the Hinode/SOT/BFI instrument 
also using observed Mercury transit images, but following the approach successfully
applied to MDI images by Mathew et al. (\cite{mathew}). We use the retrieved PSF for
the deconvolution of the images observed with the Hinode/SOT/BFI instrument
to recover the original intensities. In Sect. 2 we describe the method used
for the retrieval of the PSF. In Sect. 3 we present initial results showing
the difference in granulation contrast before and after the image correction.
Conclusions are given in Sect. 4.
\begin{figure}
\centerline{{\epsfig{file=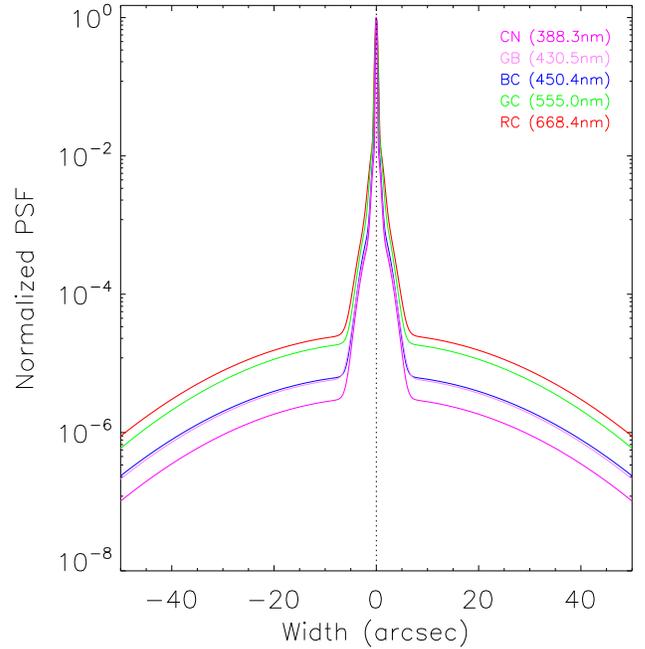,width=9.5cm}}}
\caption{PSF for different wavelengths observed by Hinode/SOT/BFI. The corresponding values for
the weights and widths of the Gaussian functions are given in table 1.}
\end{figure} 
\section{Image correction}
We used images recorded on 8th Nov. 2006 during a Mercury transit to obtain the PSF of the 
instrument. Full pixel resolution (0.05448 arcsec/pixel) images in the CN and G-bands and also in blue, 
green and red continuum bands were available. The  theoretical resolution ($\lambda / D$) 
of the Hinode/SOT ranges from 0.16 to 0.28 arc-sec between the CN-band  and red continuum wavelengths. 
In the absence of instrumental scattered light Mercury  should appear completely 
black, whereas in the observations it obviously is not (Fig. 1). 
The open circles in Fig. 1 indicate the radial (azimuthally averaged) intensity profile of Mercury 
observed in the G-band, which shows a scattered light level of around 7 - 8\% near the
center of Mercury's disk. Significant stray light is also found at all the 
other wavelengths,  with an increase in the scattered light level from shorter to 
longer wavelengths (see the additional electronic material for similar plots at 
other wavelengths; Fig. A.1).
\begin{table*}
\caption{Retrieved parameters for four-Gaussian PSFs. Width is specified in arc-secs.}
\label{table:1}
\centering
\begin{tabular}{ccccccccc}
\hline \hline
Wavelength & \multicolumn{2}{c}{Gaussian I} & \multicolumn{2}{c}{Gaussian II} & 
\multicolumn{2}{c}{Gaussian III} & \multicolumn{2}{c}{Gaussian IV}\\
(\AA) & weight & width  & weight & width & weight & width & weight & width \\
\hline
3883 (CN) & 0.6489 & 0.1084 & 0.1794 & 0.5158 & 0.1152 & 1.6999 & 0.0565 & 18.907 \\
4305 (GB) & 0.6392 & 0.1250 & 0.1721 & 0.5329 & 0.1014 & 1.8076 &  0.0873 & 19.047 \\
4504 (BC) & 0.6423 & 0.1390 & 0.1748 & 0.5212 & 0.1050 & 1.6844 &  0.0780 & 19.173 \\
5550 (GC) & 0.6250 & 0.1960 & 0.1698 & 0.6505 & 0.0920 & 1.6888 &  0.1131 & 18.686 \\
6684 (RC) & 0.6228 & 0.2188 & 0.1601 & 0.8181 & 0.0927 & 1.8330 &  0.1243 & 19.018 \\
\hline
\end{tabular}
\end{table*}

In order to obtain the PSF of the instrument we proceeded as follows. The intensity values within
the image of Mercury were replaced by zeros. These images were convolved with a guess PSF,
generated by a combination of four two-dimensional Gaussians. It was found that the use 
of four Gaussians, instead of three Gaussians and a Lorentzian as used in Mathew 
et al. (\cite{mathew}) for MDI full-disk images, gave a better fit in our case i.e. for the Mercury transit 
images. The Gaussians with appropriate weights and widths are representative of both the diffraction 
limited PSF and the non-ideal part which contributes to the scattered light in the instrument. 
By performing a two-dimensional convolution we make sure that the stray light from all sides 
contributes to the intensity within Mercury's image. The sum of the two-dimensional Gaussians 
used for the convolution is,
$$ PSF = \sum^{4}_{i=1}w_{i} c_{i} e^{-x^{2}/(2b_{i}^{2})} $$
where $ b_{i}$ is the width, $ c_{i} = 1/2 \pi b^{2}_{i} $ the normalization constant and 
$w_{i}$ the weight of the {\it i-th} Gaussian. The radial profile of the azimuthally 
averaged intensity within Mercury's outline  from the convolved images was iteratively 
fit to the similarly averaged observed intensity.
The solid line in Fig. 1 shows the best fit to the azimuthally averaged intensity profile 
through the G-band Mercury image. Figure 2  shows the normalized PSF for
different wavelengths observed with Hinode/SOT/BFI. The weights and widths of the 
two-dimensional Gaussian functions are listed in Table 1. The narrowest Gaussian in all cases
closely reproduces the theoretical resolution of the telescope, while the remaining Gaussians
with broader widths mainly account for the scattered light in the telescope. 

In the next step, the best-fit parameters were used to retrieve the stray
light-free intensities from the observed images. As pointed out by
Wedemeyer-Bh\"{o}m (\cite{bohm}) the scattering could be anistropic and the resulting
PSF could be different for observations from different fields-of-views. Since
we assume that the scattering is uniform within the observed field-of-view,
we restrict our contrast analysis to around the same part of the image, where
the Mercury transit observations were obtained, but avoid the immediate
vicinity of Mercury's image to avoid being influenced by artifacts of the
deconvolution.
\begin{figure*}
\centerline{{\epsfig{file=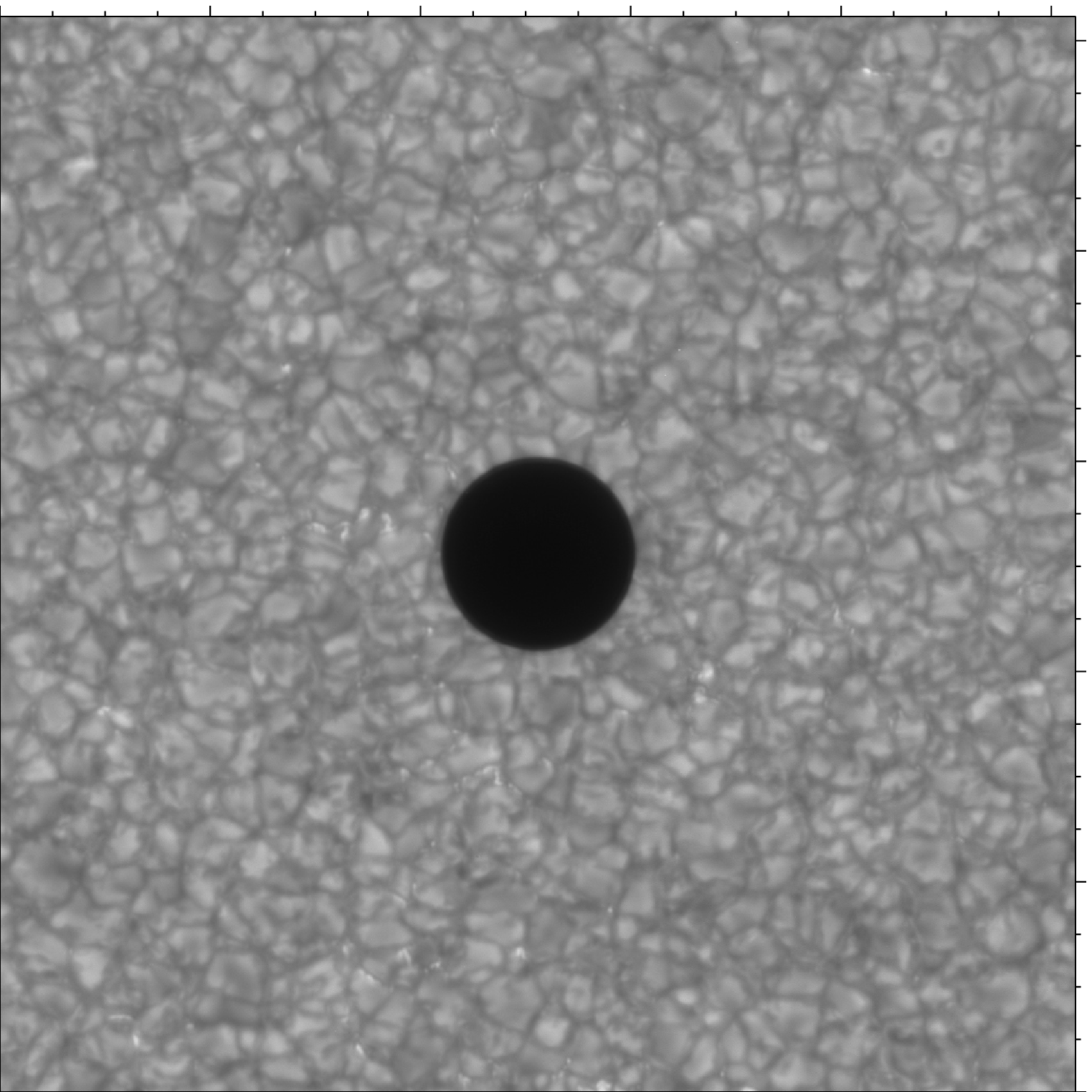,width=8.5cm}}\hspace{0.30in} {\epsfig{file=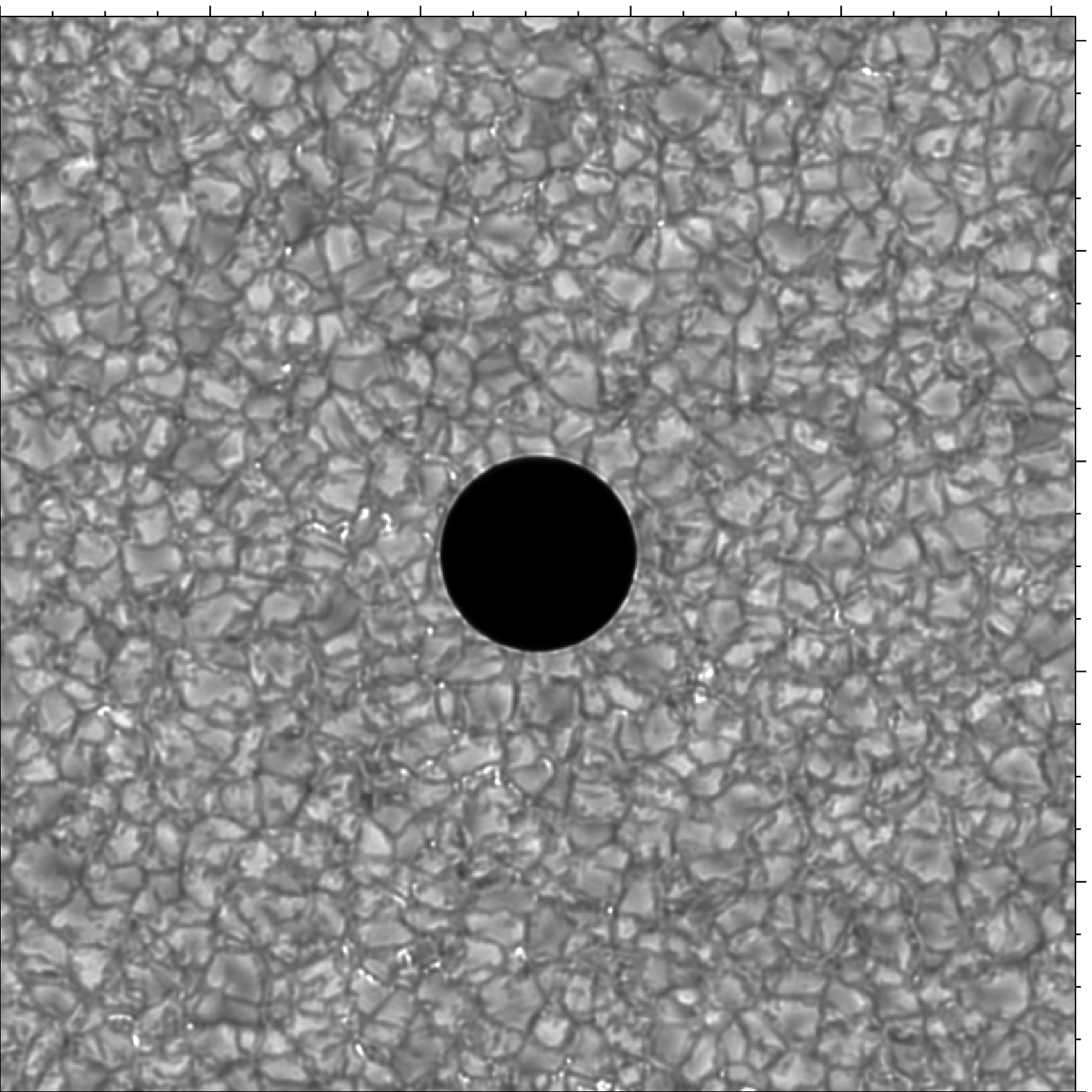,width= 8.5cm}}}
\caption{Observed ({\it left}) and corrected ({\it right}) G-band images. Each tick mark corresponds to $\sim 2.7$
arcsec. The images are cropped to $1024 \times 1024$ pixels, such that Mercury is at the center of
the FOV. Both the images are plotted on the same gray scale.}
\end{figure*}

The restoration  of the images was carried out using the 
maximum likelihood deconvolution method (an IDL routine available in Astrolib is used 
for this purpose, Richarson \cite{richardson}, Lane \cite{lane}). This method invokes an 
iterative process and updates the current estimate of the image by the product of previous 
deconvolution results and the correlation between re-convolution of the later image and the PSF. 
The convergence of restoration is checked from the chi-square of the fit which was obtained by 
comparing the observed data and the re-convolved deconvolution results.  
The filled circles in Fig. 1 show the azimuthally averaged intensity
profile across the Mercury image after the deconvolution. The intensity lies close to 
zero throughout almost the whole of Mercury's image, compared to observed values of 
7\% or higher in units of averaged quiet-Sun intensity.
This indicates that the deconvolution is effective in removing the scattered light.
In Fig. 3 an observed G-band image is shown on the left 
and the corresponding restored image on the right. The higher contrast 
of the restored image is quite marked. Note that both images have been plotted 
on the same gray-scale (for restored images in other wavelengths we refer to the additional 
electronic material to the paper; Figs. B.1 and B.2). A halo like structure is seen around the 
Mercury in the deconvolved image, which could  result from  Gibb's effect due to the large 
intensity variation across  Mercury's limb. We have investigated whether bright points are also 
affected  by comparing the observed and corrected intensity profiles across sample bright points. We have not 
observed any noticeable  signature of Gibb's effect in  bright points which could introduce errors 
in our contrast analysis.  
\section{Contrast analysis}
The deconvolution of the BFI images leads to significant enhancements of the contrast.
This is illustrated by Table 2, which compares the {\it rms} contrast of the images 
(excluding Mercury and nearby surroundings) and the intensity contrast, C, of G-band bright 
points (BPs) ($C=I/I_{0}-1$, where $I_{0}$ is the average quiet-Sun intensity) 
in each filtergram before and after deconvolution. The {\it rms} contrast is 1.6 to 2.1 times higher 
in deconvolved images than in the originally observed images. Note that since we are dealing with 
quiet-Sun data the {\it rms} contrast of the image corresponds basically to the granulation contrast, 
which lies above 20\% at two of the observed wavelengths after correction. 
We also compared the retrieved contrasts with the results obtained from 3-D
MHD simulations. By a linear fit to the contrasts obtained at different
continuum wavelengths we deduce an {\it rms} contrast of 0.128 (with a 1$\sigma$ error of
0.035) at 6302 \AA. This value is only 0.017 (i.e. 0.5 $\sigma$) lower than the
granulation contrast of 0.144 at 6302 \AA\/ obtained from 3-D radiation MHD
simulations by Danilovic et al. (\cite{danilovic}). 
  
The BP contrast given in Table 2 also increases after deconvolution of the PSF, 
by a  factor of  1.8 to 2.8. To identify the BPs we used a technique described
by Berger et al. (\cite{berger}) with an additional constraint imposed following 
Langhans et al. (\cite{langhans}). The combination of these methods secures a more reliable
BP selection (Zakharov et al. in prep.).
\begin{figure*}
\vskip 0.65cm
 \centerline{\hspace{0.25in}{\epsfig{file=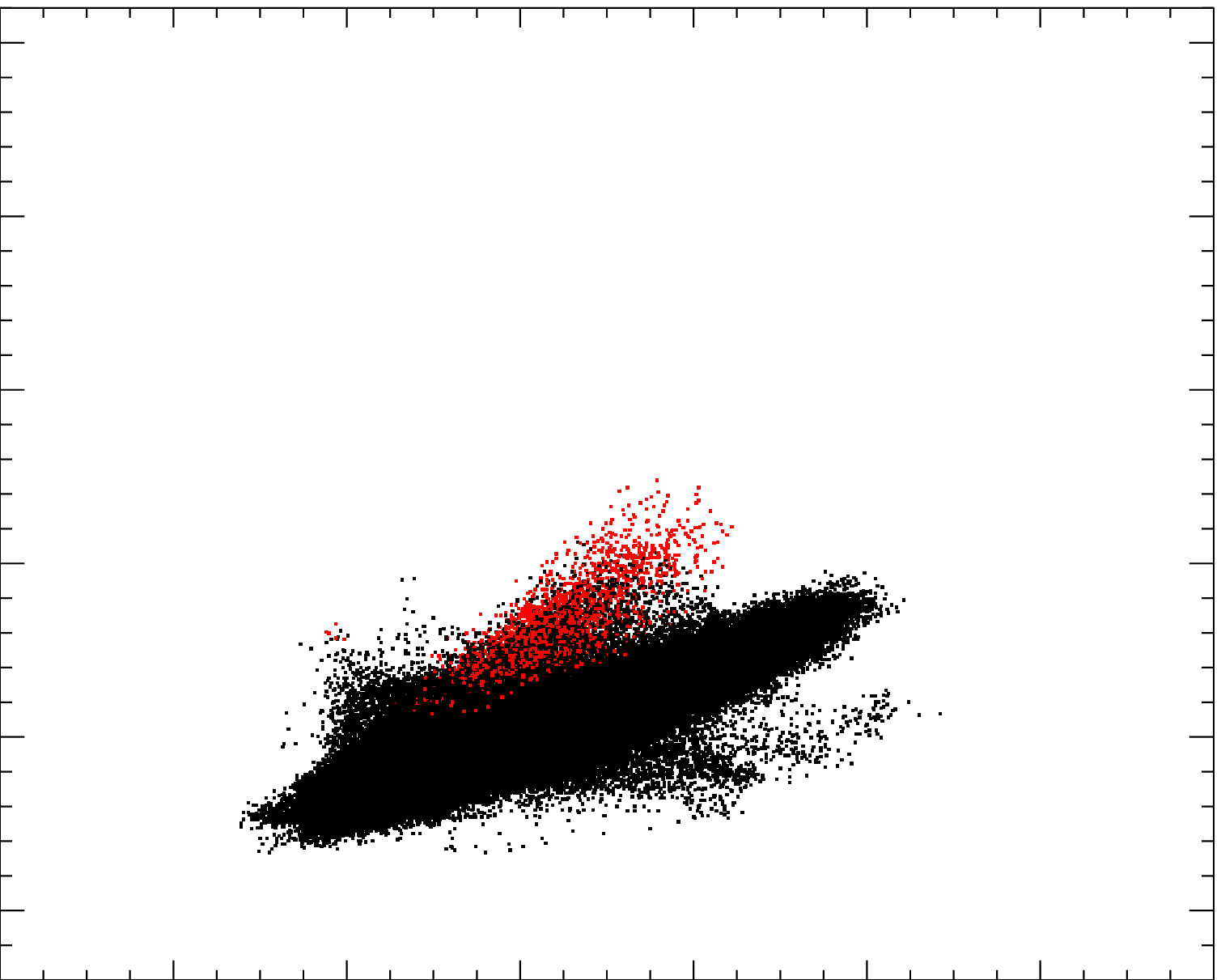,width=6.8cm}}\hspace{0.9in} {\epsfig{file=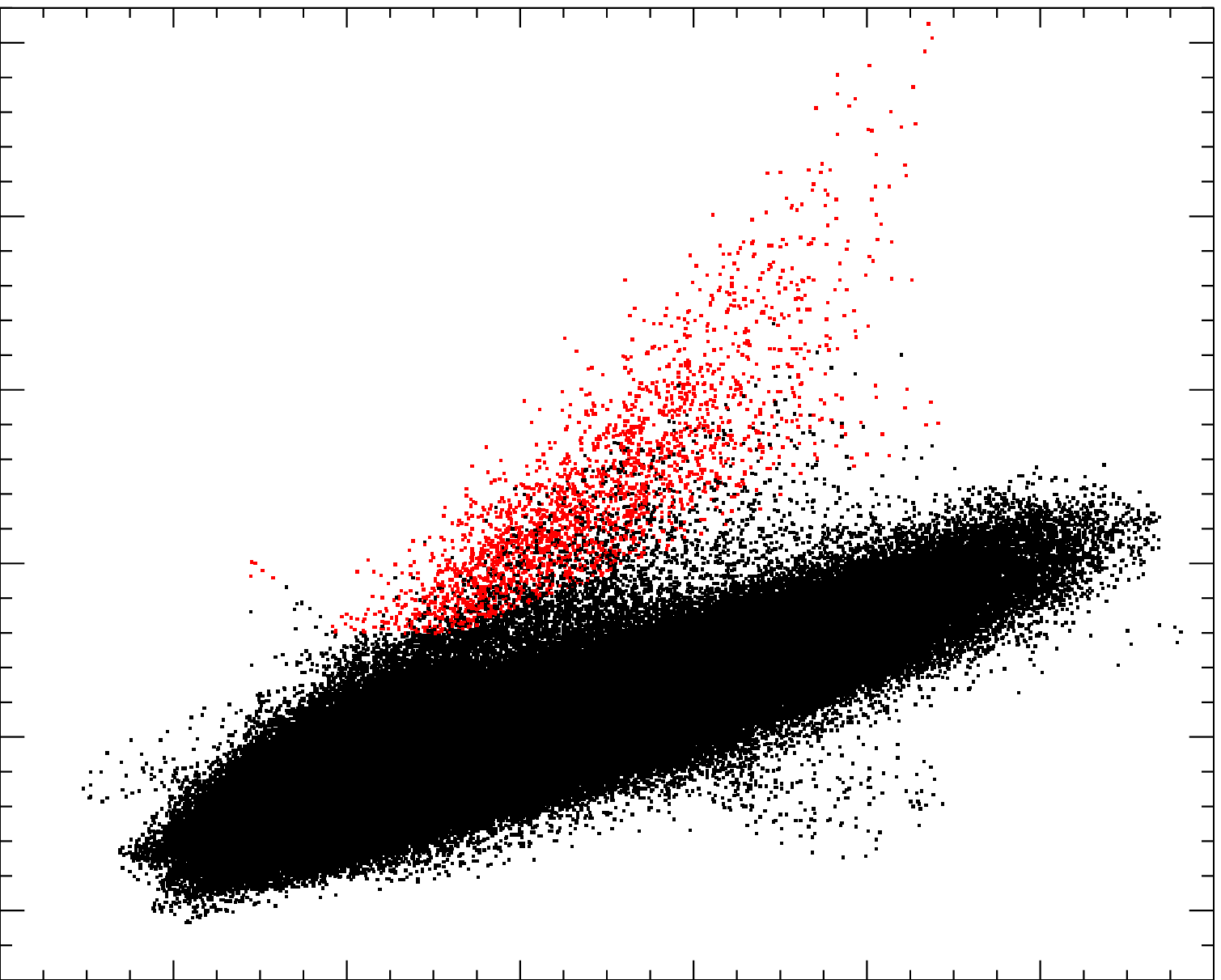,width= 6.8cm}}}
\vskip 1.0cm
 \caption{Distribution of the G-band intensity versus blue continuum intensity.
Red points correspond to BPs, black dots  to the remaining features.
All intensity values are normalized to the average intensity of that wavelength band. 
Left frame: before and right frame: after deconvolution of the PSF.}
\end{figure*}
Another way of showing the enhancement in the contrast through the deconvolution, is to plot the 
G-band intensity versus the blue continuum intensity (BC). Figure 4 displays the well-known 
(e.g. Berger et al. \cite{berger}, Shelyag et al. \cite{shelyag} and references therein) 
concentration of data points in two groups: granulation (in black) and BPs (in red). In order 
to avoid artifacts introduced by the different wavelengths and hence different spatial 
resolutions of the recorded images, we performed a spatial filtering of the acquired CN, G-band
and blue continuum images to set them to a common spatial resolution of 0.22~arcsec,
which corresponds to the diffraction limit of SOT at $\lambda=450.4~\mathrm{nm}$.
\begin{table}
\caption{{\it rms} contrast of the images and mean BP contrast values obtained before and after
deconvolution.}
\centering
\begin{tabular}{ccccc}
\hline\hline
 Wavelength & \multicolumn{2}{c}{Observed} & \multicolumn{2}{c}{Deconvolved} \\
(\AA) & rms & $\left< C\right>$ & rms & $\left< C\right>$\\
\hline
3883 (CN) & 0.134 & 0.379 & 0.218 & 0.784 \\
4305 (GB) & 0.111 & 0.293 & 0.192 & 0.702 \\
4504 (BC) & 0.122 & 0.028 & 0.215 & 0.077\\
5550 (GC) & 0.075 & 0.038 & 0.158 & 0.097\\
6684 (RC) & 0.054 & 0.038 & 0.111 & 0.069\\
\hline
\end{tabular}
\end{table}

The mean BP contrast in the CN-band is higher than in the G-band, both in the original data as 
well as in the deconvolved data by a factor of 1.29 and 1.12, respectively. 
This finding qualitatively confirms the theoretical predictions 
of Berdyugina et al. (\cite{berdyugina}) and ground-based imaging results (Rutten et al. 
\cite{rutten}, Zakharov et al. \cite{zakharov2007}). Figure 5  
displays the dependence of the ratio $C_{CN}/C_{GB}$ on the G-band intensity (after PSF 
deconvolution), where $C_{CN}$ and $C_{GB}$ are the contrasts of the respective bands as 
defined earlier in this section. In order to obtain the right slope, we bin together points 
with similar G-band intensities such that each bin contains 20 intensity values. The solid line 
represents a linear regression fit and the error bars are plotted for $\pm 1\sigma $ deviations 
due to the uncertainty in the regression gradient. From this analysis we obtain a statistically 
significant negative slope of 0.054, which indicates that brighter BPs tends to have more similar 
CN-band and G-band contrasts.  This is in qualitative agreement with the result of Zakharov et al. 
(\cite{zakharov2007}).
\begin{figure}
 \centerline{{\epsfig{file=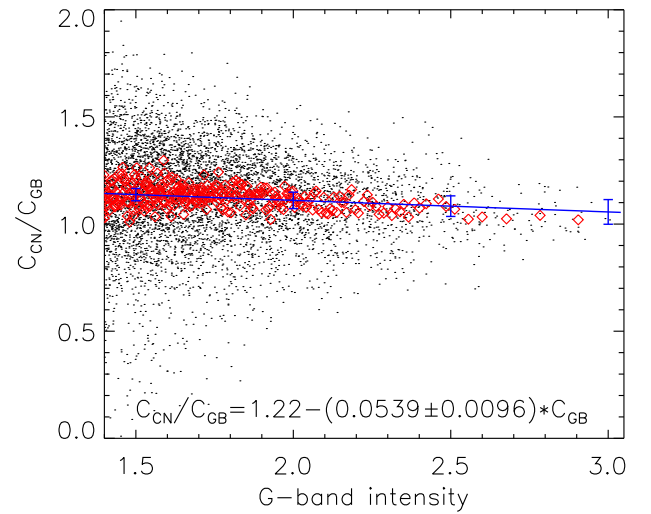,width=9.5cm}}}
 \caption{Ratio of BP contrasts in the CN-band and in the G-band versus the
 G-band intensity, normalized to its mean. Individual data points are shown as black dots and
red symbols are for the binned data. The solid blue line is a linear regression fit to the 
binned data and the error bars for $\pm 1\sigma$ uncertainty in the gradient.}
\end{figure}
\section{Conclusions}
Filtergrams recorded by the Hinode/SOT broad-band imager were used to analyze the contrast of 
the bright points and the granulation in various wavelength bands. The observed images were 
deconvolved with the retrieved PSF of the instrument in order to obtain the original 
intensities. The retrieved images are nearly free from scattered light and show a 1.6 - 2.1 times 
higher {\it rms} contrast than the untreated images, while the contrast of BPs increases by a 
factor of 1.8 - 2.8. The stronger enhancement of the BP contrast is due to their smaller size 
than granules, which are dominantly responsible for the {\it rms} contrast of the whole image.  

\acknowledgement{Hinode is a Japanese mission developed and launched by ISAS/JAXA, with NAOJ 
as domestic partner and NASA and STFC (UK) as international partners. It is operated by these 
agencies in co-operation with ESA and NSC (Norway). We would like to thank the referee for useful 
suggestions. This work has been (partially) supported by the WCU grant (No. R31-10016) funded by 
the Korean Ministry of Education, Science and Technology.}

\end{document}